\documentclass[preprint,3p]{elsarticle}
\usepackage{graphics}
\usepackage{amssymb}
\usepackage{amsthm}
\usepackage[mathlines]{lineno}
\usepackage{graphicx}
\usepackage{units}
\usepackage{url}
\usepackage{amsmath}
\usepackage{amsfonts}
\usepackage{bm}
\usepackage{textcomp}
\usepackage{subfigure}
\usepackage{multicol}
\usepackage{verbatim}
\usepackage{booktabs}
\usepackage{multirow}
\usepackage{xspace}
\usepackage[colorlinks=true,linkcolor=blue]{hyperref}

\journal{Physics Letters B}

\newcommand{\dsplus}{D^{+}_{s}}

\newcommand{\pip}{\pi^+}

\newcommand{\pim}{\pi^-}

\newcommand{\gevc}{\,\unit{GeV}/c}
\newcommand{\gevcc}{\,\unit{GeV}/c^2}
\newcommand{\br}[1]{\mathcal{B}(#1)}

\begin{document}
\begin{frontmatter}

\title{Measurement of the branching fractions of $D_{s}^{+}\rightarrow \eta'X$ and $D_{s}^{+}\rightarrow \eta'\rho^{+}$ in $e^+e^-\to D^+_{s}D^-_{s}$}

\author{
  \begin{small}
    \begin{center}
      M.~Ablikim$^{1}$, M.~N.~Achasov$^{9,f}$, X.~C.~Ai$^{1}$,
      O.~Albayrak$^{5}$, M.~Albrecht$^{4}$, D.~J.~Ambrose$^{44}$,
      A.~Amoroso$^{48A,48C}$, F.~F.~An$^{1}$, Q.~An$^{45,a}$,
      J.~Z.~Bai$^{1}$
% ...
      \\
      \vspace{0.2cm}
      (BESIII Collaboration)\\
      \vspace{0.2cm} {\it
        $^{1}$ Institute of High Energy Physics, Beijing 100049, People's Republic of China\\
% ...
$^{4}$ Bochum Ruhr-University, D-44780 Bochum, Germany\\
        $^{5}$ Carnegie Mellon University, Pittsburgh, Pennsylvania 15213, USA\\
% ...
$^{9}$ G.I. Budker Institute of Nuclear Physics SB RAS (BINP), Novosibirsk 630090, Russia\\
% ...
$^{44}$ University of Rochester, Rochester, New York 14627, USA\\
        $^{45}$ University of Science and Technology of China, Hefei 230026, People's Republic of China\\
% ...
$^{48}$ (A)University of Turin, I-10125, Turin, Italy; (B)University of Eastern Piedmont, I-15121, Alessandria, Italy; (C)INFN, I-10125, Turin, Italy\\
% ...
$^{a}$ Also at State Key Laboratory of Particle Detection and Electronics, Beijing 100049, Hefei 230026, People's Republic of China\\
% ...
$^{f}$ Also at the Novosibirsk State University, Novosibirsk, 630090, Russia\\
% ...
      } 
    \end{center}
    \vspace{0.4cm}
  \end{small}
}

\vspace{0.4cm}


\begin{abstract}
We study $D_{s}^{+}$ decays to final states involving the $\eta'$ with
a 482\,pb$^{-1}$ data sample collected at $\sqrt{s}$ = 4.009\,GeV with
the \mbox{BESIII} detector at the BEPCII collider. We measure the
branching fractions $\mathcal{B}(D^+_{s}\rightarrow \eta'X)$ =
(8.8$\pm$1.8$\pm$0.5)\%  and $\mathcal{B}(D_{s}^{+}\rightarrow
\eta'\rho^{+})$ = ($5.8\pm1.4\pm0.4$)\% where the first uncertainty is
statistical and the second is systematic.
In addition, we estimate an upper limit on the non-resonant branching
ratio $\mathcal{B}(D_{s}^{+}\rightarrow \eta'\pi^+\pi^0)<5.1\%$ at the 90\% confidence level.
Our results are consistent with CLEO's recent measurements and help to
resolve the disagreement between the theoretical prediction and CLEO's
previous measurement of $\mathcal{B}(D_{s}^{+}\rightarrow
\eta'\rho^{+})$.
\end{abstract}


\end{frontmatter}

\section{Introduction}
Hadronic weak decays of charmed mesons provide important information on flavor mixing, $CP$ violation, and strong-interaction effects~\cite{Artuso:2008vf}.
There are several proposed QCD-derived theoretical approaches to handle heavy meson decays~\cite{M.Beneke, Y.Y.Keum, C.D.Lu, C.W.Bauer, Zou:2013ez}. However, in contrast to $B$ mesons, theoretical treatment of charmed mesons suffers from large uncertainties since the $c$ quark mass is too light for good convergence of the heavy quark expansion but still much too massive for chiral perturbative theory to be applicable. Currently, theoretical results for the partial decay widths of ground-state charmed mesons agree fairly well with experimental results. However,
there exists a contradiction concerning the branching fraction $\br{\dsplus\to\eta'\rho^+}$. CLEO reported $(12.5\pm2.2)$\%~\cite{CLEO1998}, while a generalized factorization method~\cite{Fusheng:2011tw} predicts a factor of four less, \mbox{$(3.0\pm 0.5)$\%}.
Summing the large experimental value of $\br{\dsplus\to\eta'\rho^+}$ with other exclusive rates involving $\eta'$ gives
$\mathcal{B}(\dsplus\to\eta'X) = (18.6\pm2.3)\%$~\cite{PDG}, while the measured inclusive decay rate $\mathcal{B}(\dsplus\to\eta'X)$ is much lower,  $(11.7\pm1.8)$\%~\cite{Dobbs:2009aa}, where $X$ denotes all possible combinations of states.
Therefore, further experimental study of the $\eta'$ decay modes is of great importance for resolving this conflict.

Recently, CLEO reported an updated measurement of $\br{D_{s}^{+}\rightarrow\eta'\pi^{+}\pi^{0}} = (5.6\pm0.5\pm0.6)$\%~\cite{cleo:newDs}; this includes the resonant process $\eta'\rho^{+}$. This is much smaller than the previous result~\cite{CLEO1998}.
In this paper, we report the measurements of the inclusive rate $\br{D_{s}^{+}\to\eta'X}$ and the exclusive rate $\br{D_{s}^{+}\to\eta'\rho^{+}}$ at the BESIII experiment.

\section{Data Sample And Detector}
The analysis is carried out using a sample of  482\,pb$^{-1}$~\cite{Ablikim:2013pgf}
$e^+e^-$ collision data collected with the BESIII detector at the center of mass energy $\sqrt{s}$ = 4.009\,GeV.

The BESIII detector, as described in detail in Ref.~\cite{bes3-detector},
has a geometrical acceptance of 93\% of the solid angle. A small-cell helium-based main drift chamber (MDC) immersed in a 1\,T magnetic field measures the momentum of charged particles with a resolution of 0.5\% at $1\gevc$.
The electromagnetic calorimeter (EMC) detects photons with a resolution of 2.5\% (5\%) at an energy of 1\,GeV  in the barrel (end cap) region.
A time-of-flight system (TOF) assists in particle identification (PID) with a time resolution of 80\,ps (110\,ps) in the barrel (end cap) region.
Our PID methods combine the TOF information with the specific energy loss ($\mathrm{d}E/\mathrm{d}x$) measurements of charged particles in the MDC to form a likelihood $\mathcal{L}(h)(h=\pi,K)$ for each hadron ($h$) hypothesis.

A {\sc geant4}-based \cite{Agostinelli:2002hh} Monte Carlo (MC) simulation software, which
includes the geometric description of the BESIII detector and
the detector response, is used to optimize the event selection
criteria, determine the detection efficiency and estimate background contributions.
The simulation includes the beam energy spread and initial-state
radiation (ISR), implemented with {\sc kkmc}~\cite{kkmc}.
Allowing for a maximum ISR photon energy of 72\,MeV, open charm processes are simulated from $D_{s}^{+}D_{s}^{-}$ threshold at 3.937\,GeV to the center-of-mass energy 4.009\,GeV. Cross sections have been taken from Ref.~\cite{cs-cleo}.
For background contribution studies and the validation of the analysis procedure, an inclusive MC sample corresponding to an integrated luminosity of 10\,fb$^{-1}$ is analyzed.
In addition to the open charm modes, this sample includes ISR production,
continuum light quark production and QED events.
The known decay modes  are generated with {\sc evtgen}~\cite{evtgen} with branching fractions set to the world average values~\cite{PDG}, and the remaining unknown events are generated with {\sc lundcharm} \cite{lundcharm}.

\section{Data Analysis}
\subsection{Measurement of $\mathcal{B}(D_{s}^{+}\rightarrow\eta'X)$}
For data taken at 4.009\,GeV, energy conservation prohibits any additional hadrons accompanying the production of a $D_{s}^{+}D_{s}^{-}$ pair.
Following a technique first introduced by the MARK III Collaboration~\cite{mark3}, the inclusive decay rate of $D_{s}^{+}\rightarrow\eta' X$ is measured. We select single tag (ST) events in which at least one $D_s^{+}$ or $D_s^{-}$ candidate is reconstructed, and double tag (DT) events in which both $D_{s}^{+}$ and $D_{s}^{-}$ are reconstructed.
To illustrate the method, we take the ST mode $D_{s}^{-}\rightarrow\alpha$ and the signal mode $D_{s}^{+}\rightarrow\eta' X$ for example. The $\eta'$ candidates in the signal mode are reconstructed from the decay mode $\eta'\rightarrow \pi^{+}\pi^{-}\eta$ with the $\eta$ subsequently decaying into $\gamma\gamma$. The ST yields are given as
\begin{linenomath*}
\begin{equation}
y_{\rm{ST}}^{\alpha}=N_{D_{s}^{+}D_{s}^{-}} \mathcal{B}(D_{s}^{-}\rightarrow\alpha)
\varepsilon_{\rm{ST}}^{\alpha},
\end{equation}
\end{linenomath*}
where $N_{D_{s}^{+}D_{s}^{-}}$ is the number of produced $D_{s}^{+}D_{s}^{-}$ pairs and $\varepsilon_{\rm{ST}}^{\alpha}$ is the detection efficiency of reconstructing $D_{s}^{-}\rightarrow \alpha$. Similarly, the DT yields are given as
\begin{linenomath*}
\begin{equation}
y_{\rm{DT}}^{\alpha}=N_{D_{s}^{+}D_{s}^{-}}\mathcal{B}(D_{s}^{-}\rightarrow\alpha)
\mathcal{B}(D_{s}^{+}\rightarrow\eta'X)\mathcal{B}^{\rm PDG}_{\eta'}\varepsilon_{\rm{DT}}^{\alpha},
\end{equation}
\end{linenomath*}
where $\mathcal{B}^{\rm PDG}_{\eta'}$ is the product branching fractions $\mathcal{B}(\eta'\rightarrow\pi^{+}\pi^{-}\eta)\cdot\mathcal{B}(\eta\rightarrow\gamma\gamma)$, $\varepsilon_{\rm{DT}}^{\alpha}$ is the detection efficiency of reconstructing $D_{s}^{-}\rightarrow \alpha$ and $D_{s}^{+}\rightarrow \eta'X$ at the same time.
With $\varepsilon_{\rm{ST}}^{\alpha}$ and $\varepsilon_{\rm{DT}}^{\alpha}$ estimated from MC simulations, the ratio of $y_{\rm{DT}}^{\alpha}$ to
$y_{\rm{ST}}^{\alpha}$ provides a measurement of $\mathcal{B}(D_{s}^{+}\rightarrow\eta'X)$,
\begin{linenomath*}
\begin{equation}
\mathcal{B}(D_{s}^{+}\rightarrow\eta'X)\mathcal{B}^{\rm PDG}_{\eta'}=\frac{y_{\rm{DT}}^{\alpha}}{y_{\rm{ST}}^{\alpha}}
\cdot\frac{\varepsilon_{\rm{ST}}^{\alpha}}{\varepsilon_{\rm{DT}}^{\alpha}}.
\end{equation}
\end{linenomath*}
When multiple ST modes are used, the branching fraction is determined as
\begin{linenomath*}
\begin{equation}
\mathcal{B}(D_{s}^{+}\rightarrow\eta'X)\mathcal{B}^{\rm PDG}_{\eta'}
=\frac{\sum_{\alpha}y_{\rm{DT}}^{\alpha}}{\sum_{\alpha}y_{\rm{ST}}^{\alpha}
\cdot\frac{\varepsilon_{\rm{DT}}^{\alpha}}{\varepsilon_{\rm{ST}}^{\alpha}}}
=\frac{y_{\rm{DT}}}{\sum_{\alpha}y_{\rm{ST}}^{\alpha}
\cdot\frac{\varepsilon_{\rm{DT}}^{\alpha}}{\varepsilon_{\rm{ST}}^{\alpha}}},
\label{equ:sig equ}
\end{equation}
\end{linenomath*}
where $y_{\rm{DT}}=\sum_{\alpha}y_{\rm{DT}}^{\alpha}$ is the total number of DT events.

In this analysis, the ST events are selected by reconstructing a $D_{s}^{-}$ in
nine different decay modes: $K_{S}^{0}K^{-}$, $K^{+}K^{-}\pi^{-}$, $K^{+}K^{-}\pi^{-}\pi^{0}$, $K_{S}^{0}K^{+}\pi^{-}\pi^{-}$, $\pi^{+}\pi^{-}\pi^{-}$, $\pi^{-}\eta$, $\pi^{-}\eta'(\eta'\to \pi^{+}\pi^{-}\eta)$, $\pi^{-}\eta'(\eta'\to\rho^{0}\gamma, \rho^{0} \to \pi^{+}\pi^{-})$, and $\pi^{-}\pi^{0}\eta$.
The DT events are selected by further reconstructing an $\eta'$ among the remaining particles not used in the ST reconstruction.
Throughout the paper, charged-conjugate modes are always implied.

For each charged track (except for those used for reconstructing
$K_{S}^{0}$ decays), the polar angle in the MDC must
satisfy $|\cos\theta|<0.93$, and the point of closest
approach to the $e^{+}e^{-}$ interaction point (IP) must be
within $\pm$10\,cm along the beam direction and within 1\,cm
in the plane perpendicular to the beam direction.
A charged $K(\pi)$ meson is identified by requiring the PID likelihood to satisfy $\mathcal{L}(K) > \mathcal{L}(\pi)$ ($\mathcal{L}(\pi) > \mathcal{L}(K)$).

Showers identified as photon candidates must satisfy the following requirements. The deposited energy in the EMC is required to be larger than 25\,MeV in the barrel region ($|\cos\theta|<0.8$) or larger than 50\,MeV in the end cap region ($0.86<|\cos\theta|<0.92$). To suppress electronic noise and energy deposits unrelated to the event, the EMC time deviation from the event start time is required to be $0 \leq T \leq 700$\,ns.
Photon candidates must be separated by at least 10 degrees from the extrapolated positions of any charged tracks in the EMC.

The $K_{S}^{0}$ candidates are formed from pairs of oppositely charged tracks. For these two tracks, the polar angles in the MDC must satisfy $|\cos\theta|<0.93$, and the point of closest approach to the IP must be within $\pm$20\,cm along the beam direction. No requirements on the distance of closest approach in the transverse plane or on particle identification criteria are applied to the tracks.
Their invariant mass is required to satisfy $0.487<M(\pi^{+}\pi^{-})<0.511\gevcc$. The two tracks are constrained to originate from a common decay vertex, which is required to be separated from the IP by a decay length of at least twice the vertex resolution.

The $\pi^{0}$ and $\eta$ candidates are reconstructed from photon pairs.
The invariant mass is required to satisfy $0.115<M(\gamma\gamma)<0.150\gevcc$ for $\pi^{0}$, and $0.510<M(\gamma\gamma)<0.570\gevcc$ for $\eta$. To improve the mass resolution, a mass-constrained fit to the nominal mass of $\pi^{0}$ or $\eta$~\cite{PDG} is applied to the photon pairs.
For $\eta'$ candidates, the invariant mass must satisfy $0.943<M(\eta'_{\pi\pi\eta})<0.973\gevcc$ and $0.932<M(\eta'_{\rho\gamma})<0.980\gevcc$.
For the $\eta'_{\rho\gamma}$ candidates, we additionally require $0.570<M(\pi^{+}\pi^{-})<0.970\gevcc$ to reduce contributions from combinatorial background.

We define the energy difference, $\Delta E\equiv E-E_{0}$, where $E$ is the total measured energy of
the particles in the $D_{s}^{-}$ candidate and $E_{0}$ is the beam energy.
The $D_{s}^{-}$ candidates are rejected if they fail to pass $\Delta E$ requirements corresponding to 3 times the resolution, as given in Table~\ref{tab:deE-cut}. To reduce systematic uncertainty, we apply different requirements on $\Delta E$ for data and MC samples. If there is more than one $D_s^{-}$ candidate in a specific ST mode, the candidate with the smallest $|\Delta E|$ is kept for further analysis.

\begin{table}[tp!]
\centering
\caption{Requirements on $\Delta E$ for ST $D_{s}^{-}$ candidates.}
\begin{tabular}{lcc}
\hline \hline
ST mode $\alpha$  & data (GeV) & MC (GeV)  \\
\hline
$K_{S}^{0}K^{-}$
& $(-0.027, 0.021)$  &  $(-0.025, 0.021)$ \\
$K^{+}K^{-}\pi^{-}$
& $(-0.032, 0.023)$  &  $(-0.031, 0.024)$ \\
$K^{+}K^{-}\pi^{-}\pi^{0}$
& $(-0.041, 0.022)$  & $(-0.041, 0.022)$ \\
$K_{S}^{0}K^{+}\pi^{-}\pi^{-}$
& $(-0.035, 0.024)$  & $(-0.032, 0.026)$ \\
$\pi^{+}\pi^{-}\pi^{-}$
& $(-0.036, 0.023)$  & $(-0.033, 0.025)$ \\
$\pi^{-}\eta$
&  $(-0.038, 0.037)$ & $(-0.041, 0.032)$ \\
$\pi^{-}\eta'_{\pi\pi\eta}$
&  $(-0.035, 0.027)$ & $(-0.034, 0.028)$ \\
$\pi^{-}\eta'_{\rho\gamma}$
&  $(-0.035, 0.022)$ &  $(-0.035, 0.021)$ \\
$\pi^{-}\pi^{0}\eta$
& $(-0.053, 0.030)$  & $(-0.053, 0.028)$ \\
\hline \hline
\end{tabular}
\label{tab:deE-cut}
\end{table}

To identify ST signals, the beam-constrained mass $M_{\rm{BC}}$ is used.
This is the mass of the $D_{s}^{-}$ candidate calculated by substituting the beam energy $E_{0}$ for the measured energy of the $D_{s}^{-}$ candidate: $M_{\rm{BC}}^{2}c^{4} \equiv E_{0}^{2} - p^{2}c^{2}$, where $p$ is the measured momentum of the $D_{s}^{-}$ \rm candidate. True $D_{s}^{-}\to\alpha$ single-tags peak at the nominal $D_{s}^{-}$ mass in $M_{\rm{BC}}$.

\begin{figure}[tp!]
\begin{center}
\includegraphics[width=\linewidth]{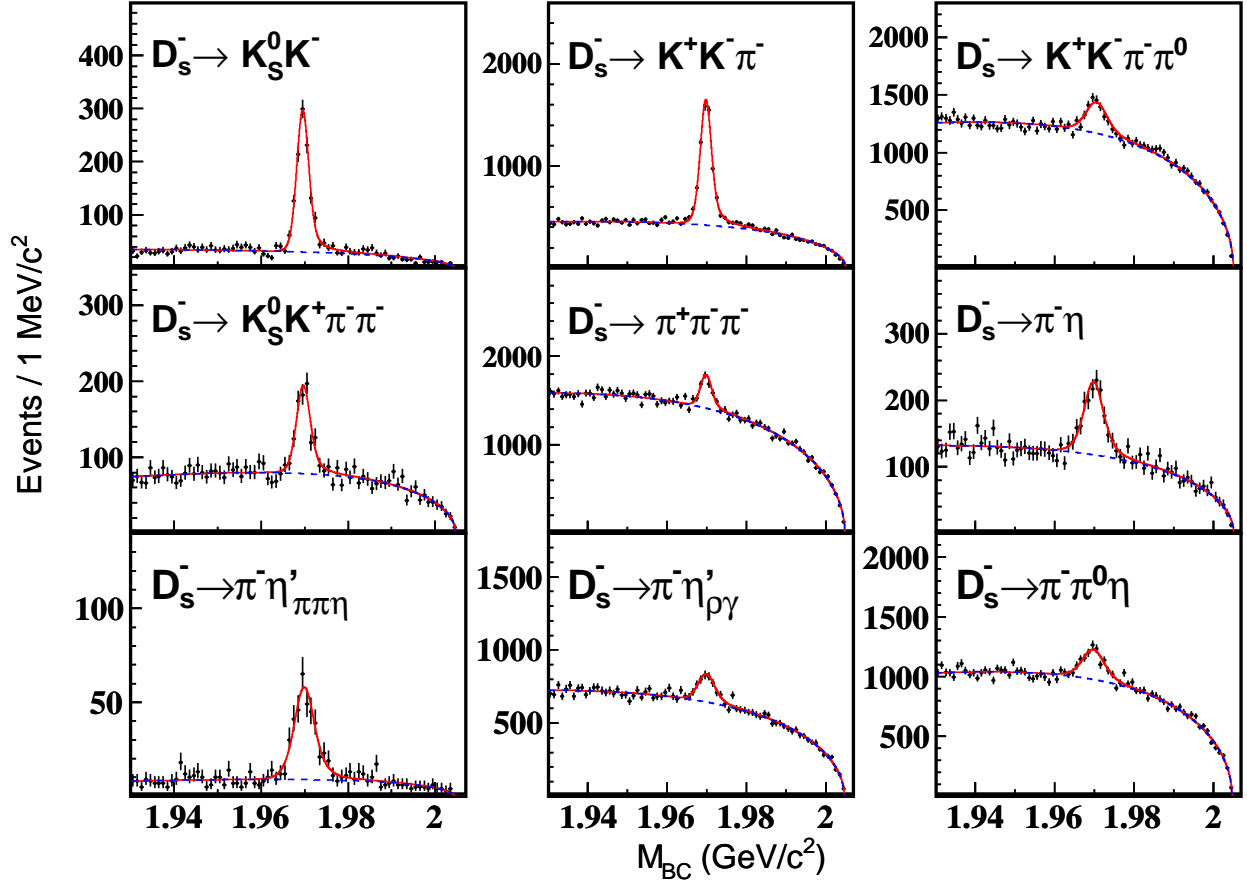}
\caption{Fits to the $M_{\rm{BC}}$ distributions for the ST $D_{s}$ candidates. In each plot, the points with error bars are data, the dashed curve is the background contribution and the solid line shows the total fit.}
\label{ST-yields}
\end{center}
\end{figure}

We fit the $M_{\rm{BC}}$ distribution of each mode $\alpha$ to obtain $y^\alpha_{\rm{ST}}$. Background contributions for each mode are well described by the ARGUS function~\cite{argus}, as verified with MC simulations.
The signal distributions are modeled by a MC-derived signal shape convoluted with a Gaussian function whose parameters are left free in the fit. The Gaussian function compensates the resolution difference between data and MC simulation.
Figure~\ref{ST-yields} shows the fits to the $M_{\rm{BC}}$ distributions in data; the fitted ST yields are presented in Table~\ref{table:STandDT} along with the detection efficiencies estimated based on MC simulations.

To select events where the $D_{s}^{+}$ decays to $\eta' X$, we require that the DT events contain an $\eta'$ candidate among the particles recoiling against the ST candidate. As mentioned above, the $\eta'$ candidates are reconstructed in the decay $\eta'\rightarrow \pi^{+}\pi^{-}\eta$, with the $\eta$ subsequently decaying into $\gamma\gamma$. All particles used in the $\eta'$ reconstruction must satisfy the requirements detailed above. If there is more than one $\eta'$ candidate, the one with the smallest $\Delta M \equiv |M(\eta'_{\pi\pi\eta})-m(\eta')|$ is kept, where $m(\eta')$ is the nominal $\eta'$ mass~\cite{PDG}. The decay mode $\eta'\rightarrow \rho^{0}\gamma$ is not used due to large contributions from combinatorial background.

\begin{figure}[tp!]
\begin{center}
\includegraphics[width=0.48\linewidth]{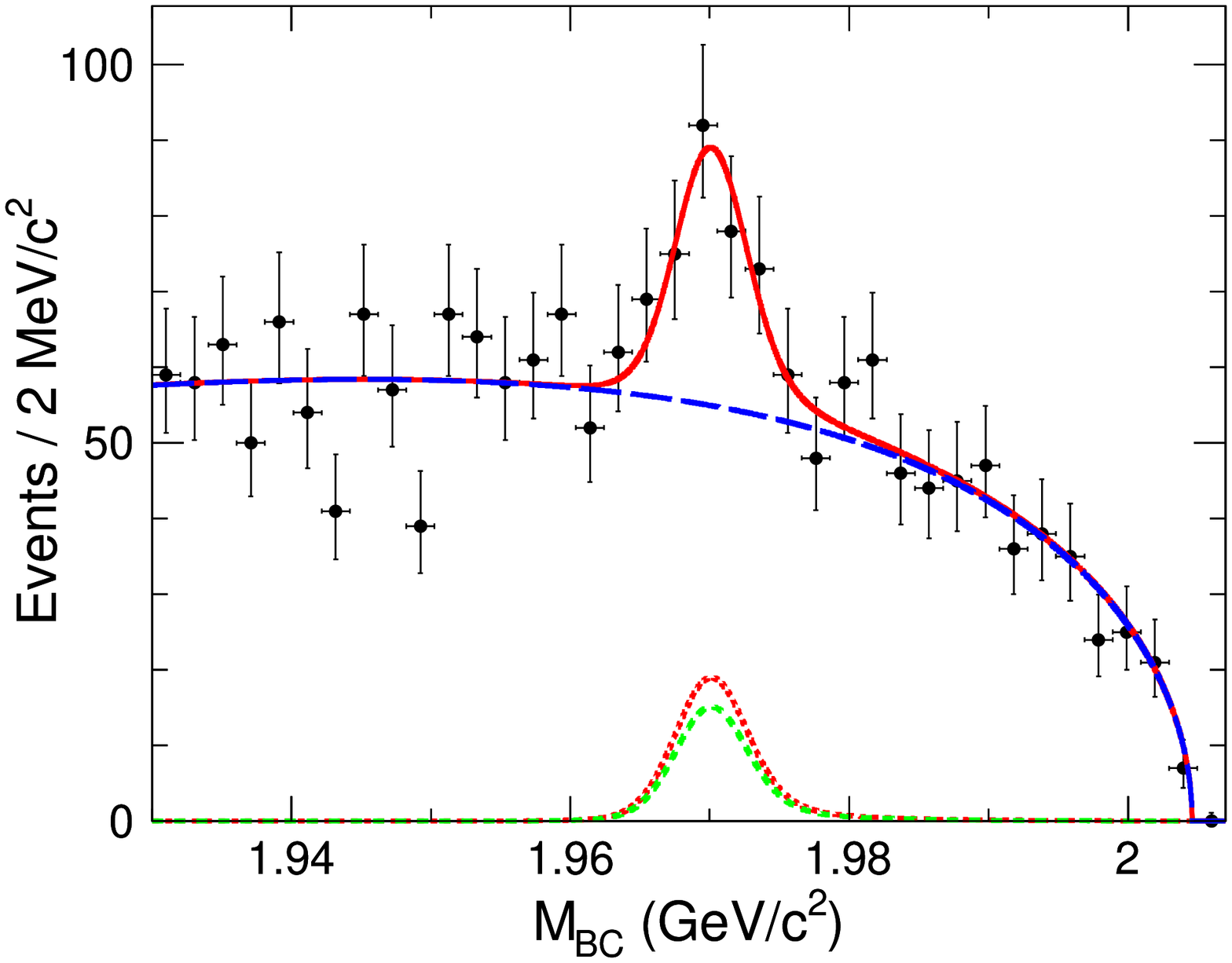}
\includegraphics[width=0.48\linewidth]{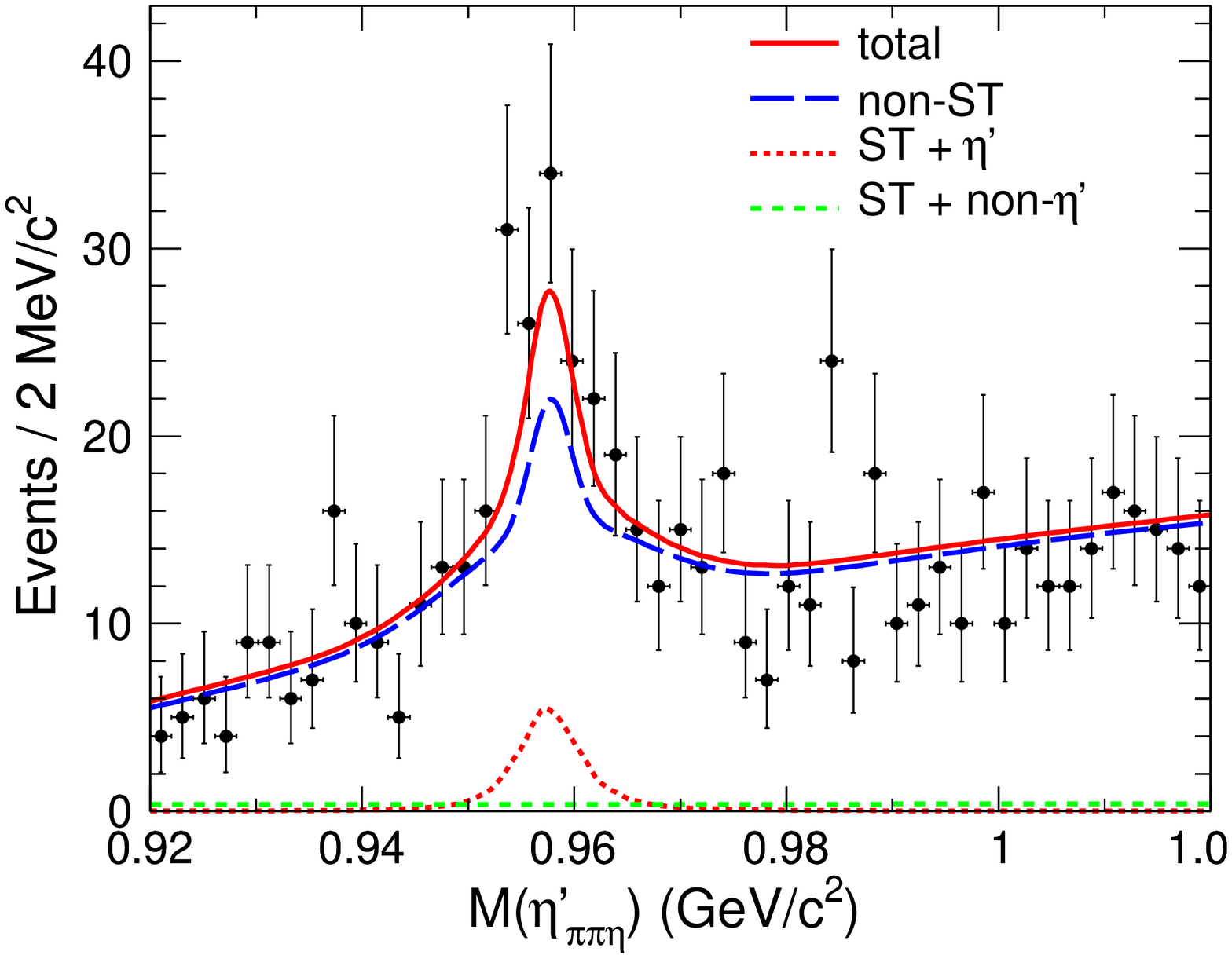}
\caption{Projections of the two-dimensional unbinned fit to DT events from data onto $M_{\rm{BC}}$ (left) and $M(\eta'_{\pi^{+}\pi^{-}\eta})$ (right).}
\label{fig:epx-2D-fit}
\end{center}
\end{figure}

There are peaking background contributions in $M(\eta'_{\pi\pi\eta})$ produced by events in which there is a wrongly-reconstructed $D_{s}^{-}$ tag accompanied by a real $\eta'$ in the rest of the event. To obtain the DT yields, we therefore perform a two-dimensional unbinned fit to the variables $M_{\rm{BC}}(\alpha)$ and $M(\eta'_{\pi\pi\eta})$.
For $M_{\rm{BC}}(\alpha)$, the fit functions are the same as those used in the extraction of $y^\alpha_{\rm{ST}}$.
For $M(\eta'_{\pi\pi\eta})$, the signal is described by the convolution of a  MC-derived signal shape and a Gaussian function with parameters left free in the fit.
Background contributions in $M(\eta'_{\pi\pi\eta})$ consist of ($a$) $D_{s}^{+}D_{s}^{-}$ events in which $D_{s}^{-}$ decays to the desired ST modes, but the $D_{s}^{+}$ decay does not involve an $\eta'$; ($b$) other (non-ST signal) decays of $D_{s}^{-}$ and also non-$D_{s}^{+}D_{s}^{-}$ processes. Component ($a$) is described with a first-order polynomial function. Component ($b$) is modeled with the sum of two Gaussian functions plus a quadratic polynomial function.
The means of the two Gaussians are fixed to the $\eta'$ nominal mass~\cite{PDG}. Other parameters and all the amplitudes are left free in the fit. The ARGUS function of $M_{\rm{BC}}(\alpha)$ helps to constrain the description of $M(\eta'_{\pi\pi\eta})$ in component ($b$). This treatment on background contributions has been verified in MC simulations.
There is no obvious correlation between $M_{\rm{BC}}(\alpha)$ and $M(\eta'_{\pi\pi\eta})$, so the probability density functions (PDFs) of these two variables are directly multiplied.
We obtain the combined DT yield $y_{\rm{DT}}$ from the unbinned fit shown in Fig.~\ref{fig:epx-2D-fit}.
Table~\ref{table:STandDT} gives the total yields of DT in data and the corresponding DT efficiencies. Combining the yields and efficiencies, we obtain $\mathcal{B}(D_{s}^{+}\rightarrow\eta'X) = (8.8\pm1.8)\%$ with Eq.~\ref{equ:sig equ}.

\begin{table}[tp!]
\centering
\footnotesize
\caption{The detection efficiencies and the data yields of the ST and DT events. The efficiencies do not include the intermediate branching fractions for $\pi^{0}\rightarrow\gamma\gamma$, $\eta\rightarrow\gamma\gamma$, $K_{S}^{0}\rightarrow\pi^{+}\pi^{-}$, $\eta'\rightarrow\pi^{+}\pi^{-}\eta$ and $\eta'\rightarrow\rho^{0}\gamma$. All uncertainties are statistical only.}
\begin{tabular}{l|c|r@{ }l|r@{ }l|c}
\hline \hline
ST mode $\alpha$     & $\varepsilon^\alpha_{\rm{ST}}$(\%)   &  \multicolumn{2}{c|}{$y^\alpha_{\rm{ST}}$}  &  \multicolumn{2}{c|}{$\varepsilon^\alpha_{\rm{DT}}$(\%)}   &   $y_{\rm{DT}}$\\
\hline
$K_{S}^{0}K^{-}$                                     & 47.89 $\pm$ 0.35  &  1088 $\pm$ &40  & 13.75 $\pm$ &0.14 &\\
$K^{+}K^{-}\pi^{-}$                                  & 44.16 $\pm$ 0.18  &  5355 $\pm$ &118 & 12.46 $\pm$ &0.14 &\\
$K^{+}K^{-}\pi^{-}\pi^{0}$                           & 13.25 $\pm$ 0.22  &  1972 $\pm$ &145 &  4.32 $\pm$ &0.08 &\\
$K_{S}^{0}K^{+}\pi^{-}\pi^{-}$                       & 24.27 $\pm$ 0.37  &   595 $\pm$ &50  &  6.05 $\pm$ &0.09 &\\
$\pi^{+}\pi^{-}\pi^{-}$                              & 60.26 $\pm$ 0.90  &  1657 $\pm$ &143 & 17.18 $\pm$ &0.16 &68$\pm$14\\
$\pi^{-}\eta$                                        & 48.39 $\pm$ 0.70  &   843 $\pm$ &54  & 14.82 $\pm$ &0.16 &\\
$\pi^{-}\eta'_{\pi\pi\eta}$                          & 29.48 $\pm$ 0.52  &   461 $\pm$ &41  &  7.91 $\pm$ &0.11 &\\
$\pi^{-}\eta'_{\rho\gamma}$                          & 43.11 $\pm$ 0.88  &  1424 $\pm$ &147 & 11.96 $\pm$ &0.13 &\\
$\pi^{-}\pi^{0}\eta$                                 & 26.02 $\pm$ 0.32  &  2260 $\pm$ &156 &  7.90 $\pm$ &0.11 &\\
\hline \hline
\end{tabular}
\label{table:STandDT}
\end{table}

\subsection{Measurement of $\mathcal{B}(D_{s}^{+}\rightarrow\eta'\rho^{+})$}
In order to improve the statistical precision, we determine the branching fraction for $D_{s}^{+}\rightarrow\eta'\rho^{+}$ using STs.
As a standalone measurement, this does not benefit from cancellation of systematic uncertainties as in the double-tag
method. However, a similar cancellation can be achieved by measuring the signal relative to a similar, already well-measured final state.
Thus, we measure $\br{D_{s}^{+}\rightarrow\eta'\rho^{+}}$ relative to $\br{D_s^{+}\rightarrow K^{+}K^{-}\pi^{+}}$, using
\begin{linenomath*}
\begin{equation}
\frac{\mathcal{B}(D_{s}^{+}\rightarrow\eta'\rho^{+}) \mathcal{B}^{\rm PDG}_{\rho^+}\mathcal{B}^{\rm PDG}_{\eta'}}{\mathcal{B}(D_{s}^{+}\rightarrow K^{+}K^{-}\pi^{+})}=\frac{y^{\eta'\rho^{+}}_{\rm{ST}}}{y^{K^{+}K^{-}\pi^{+}}_{\rm{ST}}}\cdot
\frac{\varepsilon^{K^{+}K^{-}\pi^{+}}_{\rm{ST}}}{\varepsilon^{\eta'\rho^{+}}_{\rm{ST}}},
\label{equ:eprho equ}
\end{equation}
\end{linenomath*}
where $\mathcal{B}^{\rm PDG}_{\rho^+}=\mathcal{B}(\rho^+\to\pi^+\pi^0)\mathcal{B}(\pi^0\rightarrow\gamma\gamma)$.

\begin{figure}[tp!]
\begin{center}
\includegraphics[width=0.48\linewidth]{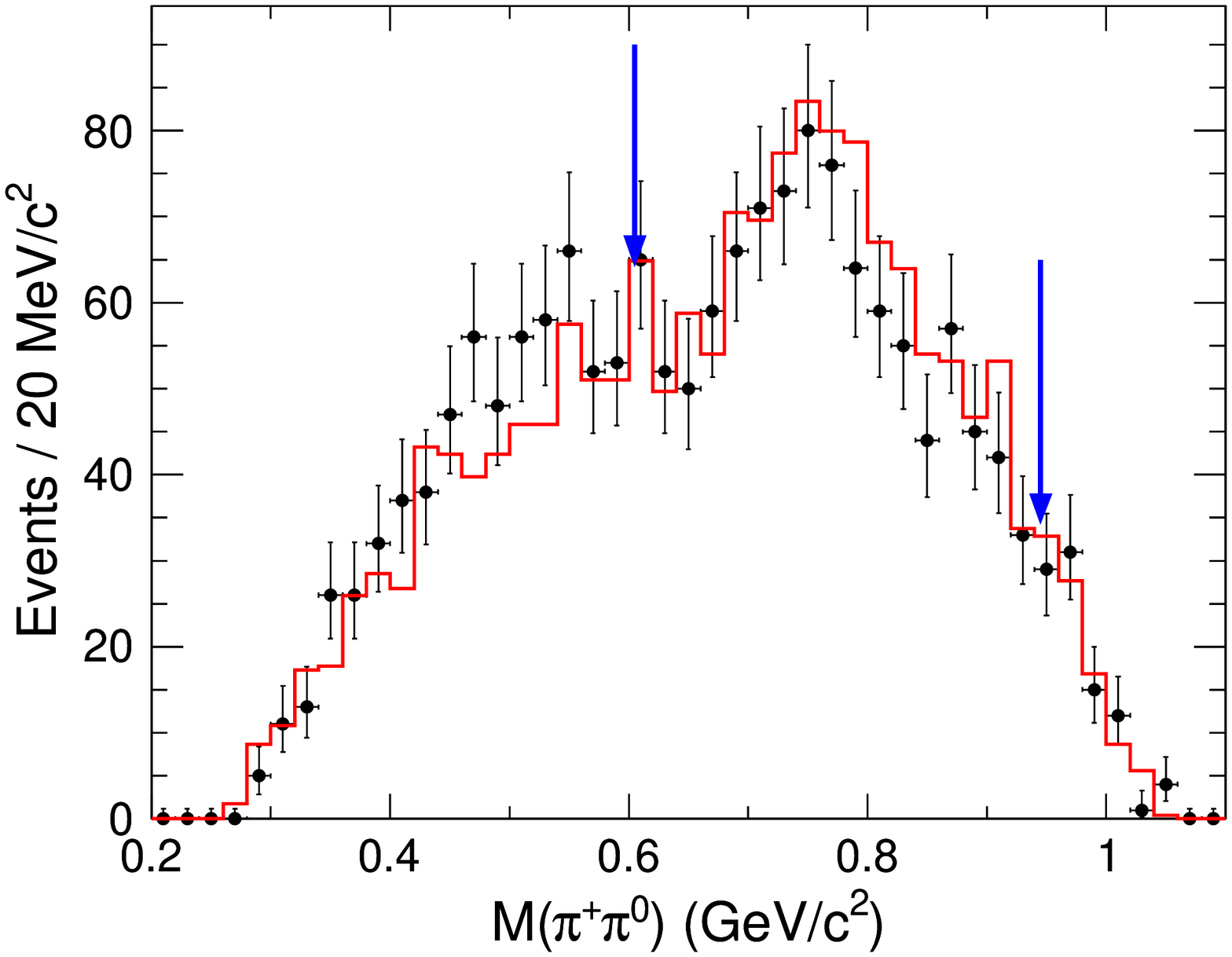}
\includegraphics[width=0.48\linewidth]{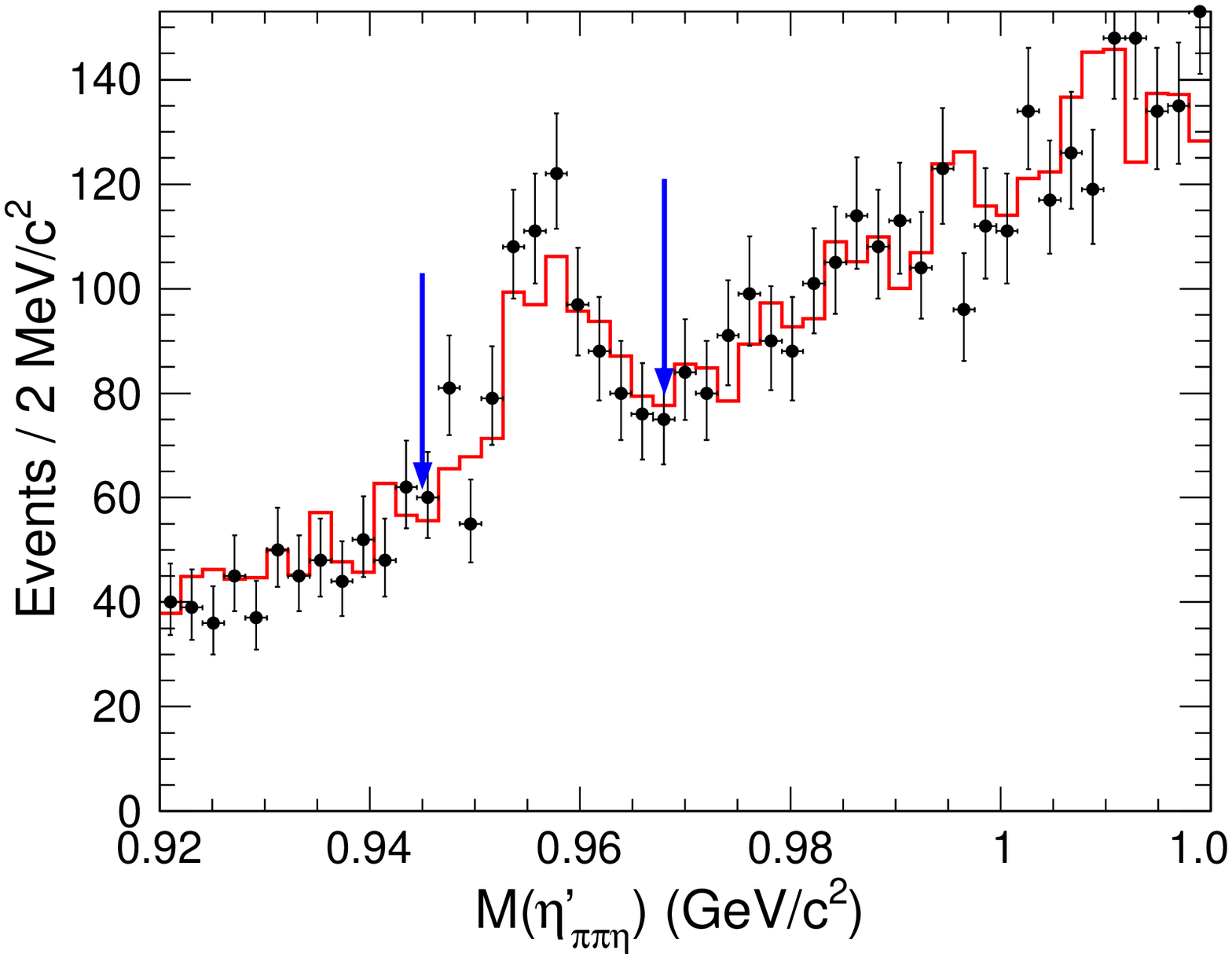}
\caption{Comparison of the $M(\pi^+\pi^0)$ (left) and $M(\eta'_{\pi\pi\eta})$ (right) distributions in ST events of $D_{s}^{+}\rightarrow\eta'\rho^{+}$ in data (points) and inclusive MC (solid line). The arrows show the signal region.}
\label{fig:rho-ep-mass-comp}
\end{center}
\end{figure}

The decay $D_s^{+}\rightarrow K^{+}K^{-}\pi^{+}$ is reconstructed in the same manner as reported above in the ST mode. Our MC simulation of this
mode includes a full treatment of interfering resonances in the Dalitz plot~\cite{DalitzKKpi}.
The decay $D_{s}^{+}\rightarrow\eta'\rho^{+}$ is reconstructed via the decays $\eta'\rightarrow\pi^{+}\pi^{-}\eta$ and $\rho^+\to\pi^+\pi^0$, where $\eta\;(\pi^0)\rightarrow\gamma\gamma$. We apply the same criteria to find $\pi^{0}$ and $\eta$ candidates as were used in the analysis of $D_{s}^{+}\rightarrow\eta'X$.
We do not require PID criteria on the charged tracks, but instead assume them all to be pions. In the reconstruction of $\rho^{+}$ and $\eta'$, the $\pi^{+}$ are randomly assigned. 
The invariant mass, $M(\pi^{+}\pi^{0})$, of the $\rho^{+}$ candidate is required to be within $\pm$0.170$\gevcc$ of the nominal $\rho^{+}$ mass, and the invariant mass of the $\eta'$ candidate, $M(\eta'_{\pi\pi\eta})$, is required to lie in the interval (0.943, 0.973)$\gevcc$.
Additionally requiring $1.955<M_{\rm{BC}}<1.985\gevcc$ to enrich signal events, the $M(\pi^{+}\pi^{0})$ distribution of $D_{s}^{+}\rightarrow\eta'\rho^{+}$ in inclusive MC simulations and data in Fig.~\ref{fig:rho-ep-mass-comp} show good agreement.
The small difference visible in the $M(\eta'_{\pi\pi\eta})$ distribution will be taken into account in the systematic uncertainties.

If multiple $\eta'\rho^+$ candidates are found in an event, only the one with the smallest $|\Delta E|$ is kept. We require $-0.035<\Delta E<0.023$\,GeV for data and $-0.037<\Delta E< 0.029$\,GeV for MC. Fits to the $M_{\rm{BC}}$ distributions are used to extract signal yields.
To separate the three body process \mbox{$D_{s}^{+}\rightarrow\eta'\pi^{+}\pi^{0}$} from the two body decay $D_{s}^{+}\rightarrow\eta'\rho^{+}$, the helicity angle $\theta_{\pi^{+}}$ is used to extract the $\rho^{+}$ component, where $\theta_{\pi^{+}}$ is the angle between the momentum of the $\pi^{+}$  from the $\rho^{+}$ decay and the direction opposite to the $D_{s}^{+}$ momentum in the $\rho^{+}$ rest frame.
The signal $D_{s}^{+}\rightarrow\eta'\rho^{+}$ is distributed as $\cos^{2}\theta_{\pi^{+}}$,  while the three body process is flat in $\cos\theta_{\pi^+}$.

\begin{figure}[tp!]
\begin{center}
\includegraphics[width=0.48\linewidth]{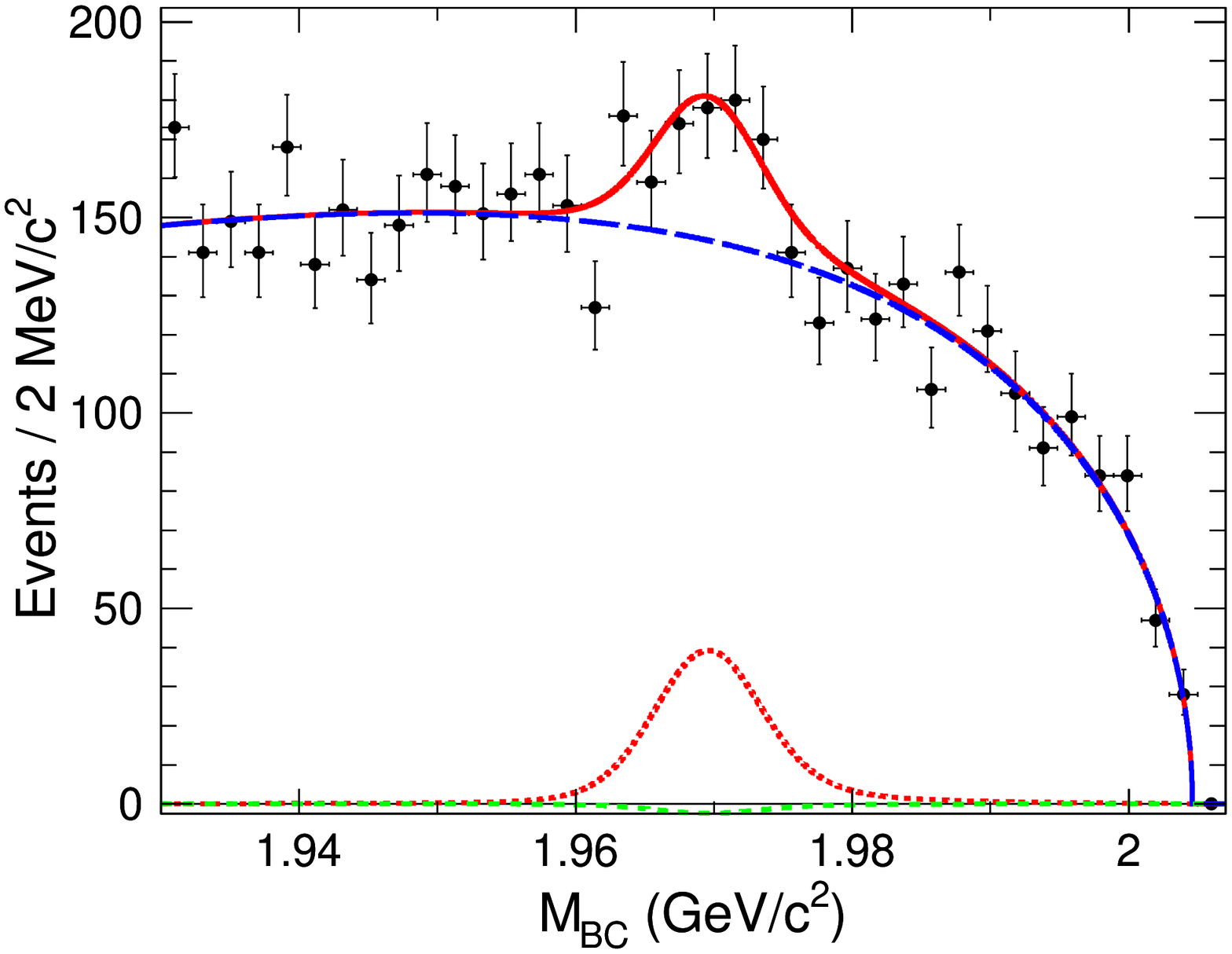}
\includegraphics[width=0.48\linewidth]{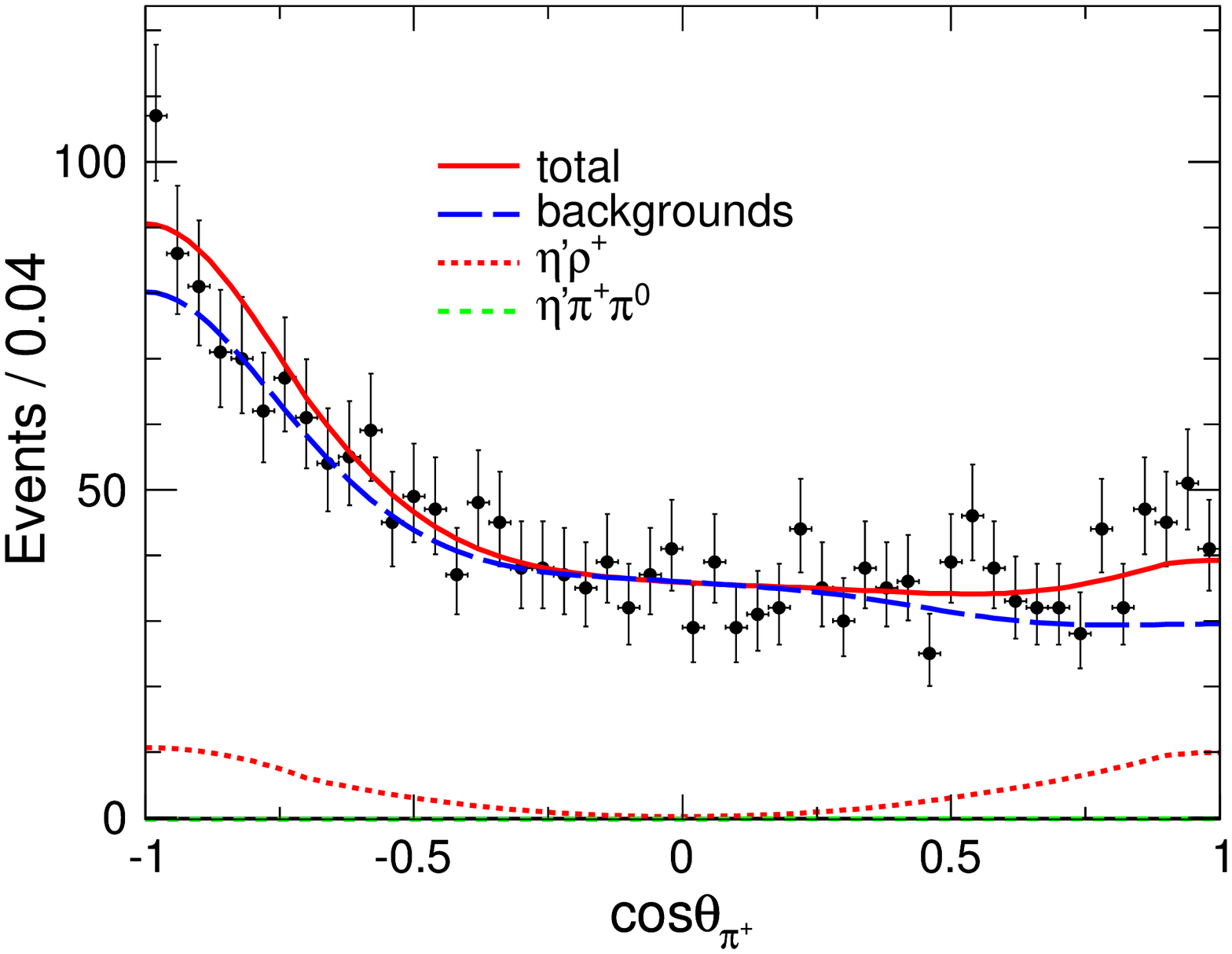}
\caption{Projection plots of the two dimensional unbinned fit onto $M_{\rm{BC}}$ (left) and $\cos\theta_{\pi^{+}}$ (right). The signal events are enriched by requiring $1.955<M_{\rm{BC}}<1.985\gevcc$ in the right plot.}
\label{fig:mbcVSctht}
\end{center}
\end{figure}

\begin{figure}[tp!]
\begin{center}
\includegraphics[width=0.48\linewidth]{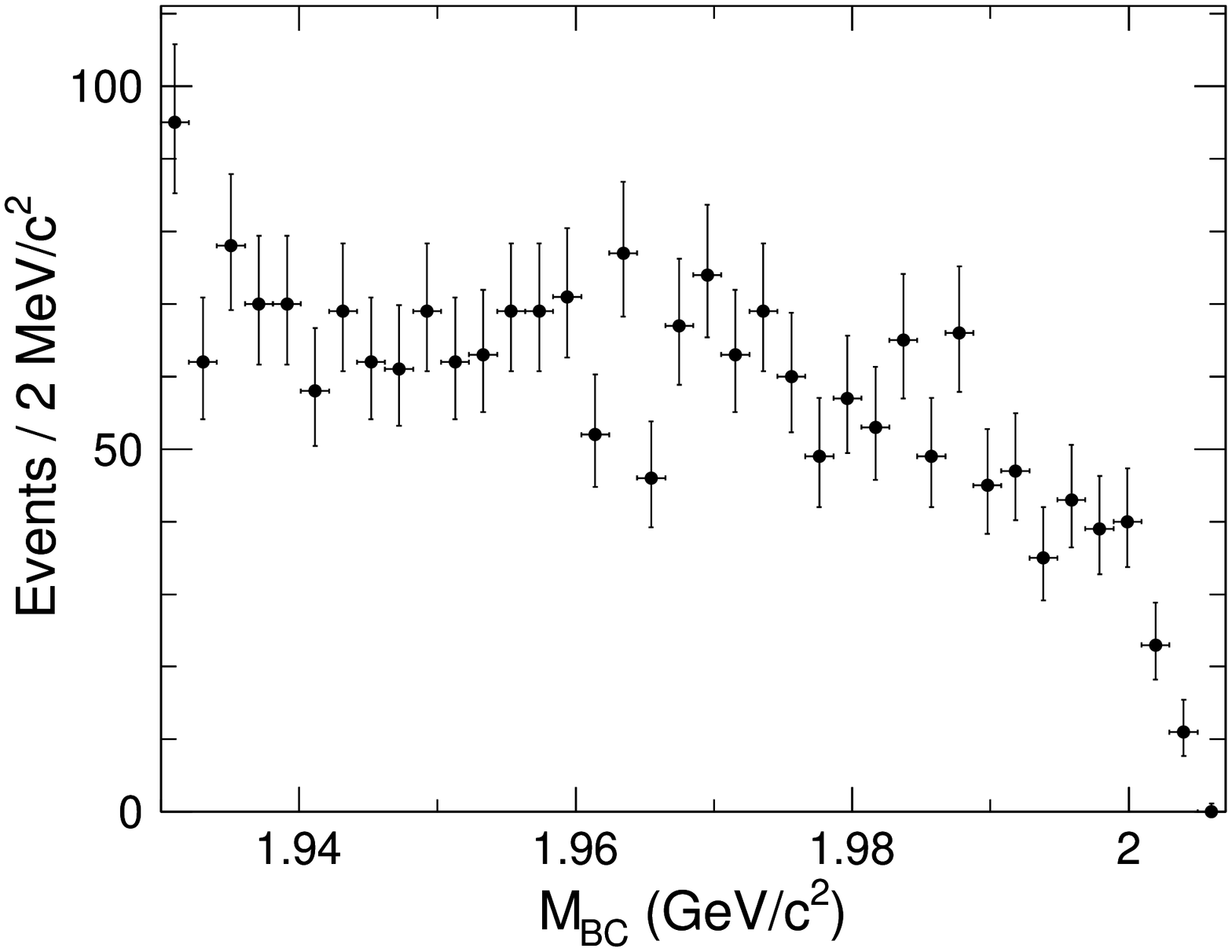}
\includegraphics[width=0.48\linewidth]{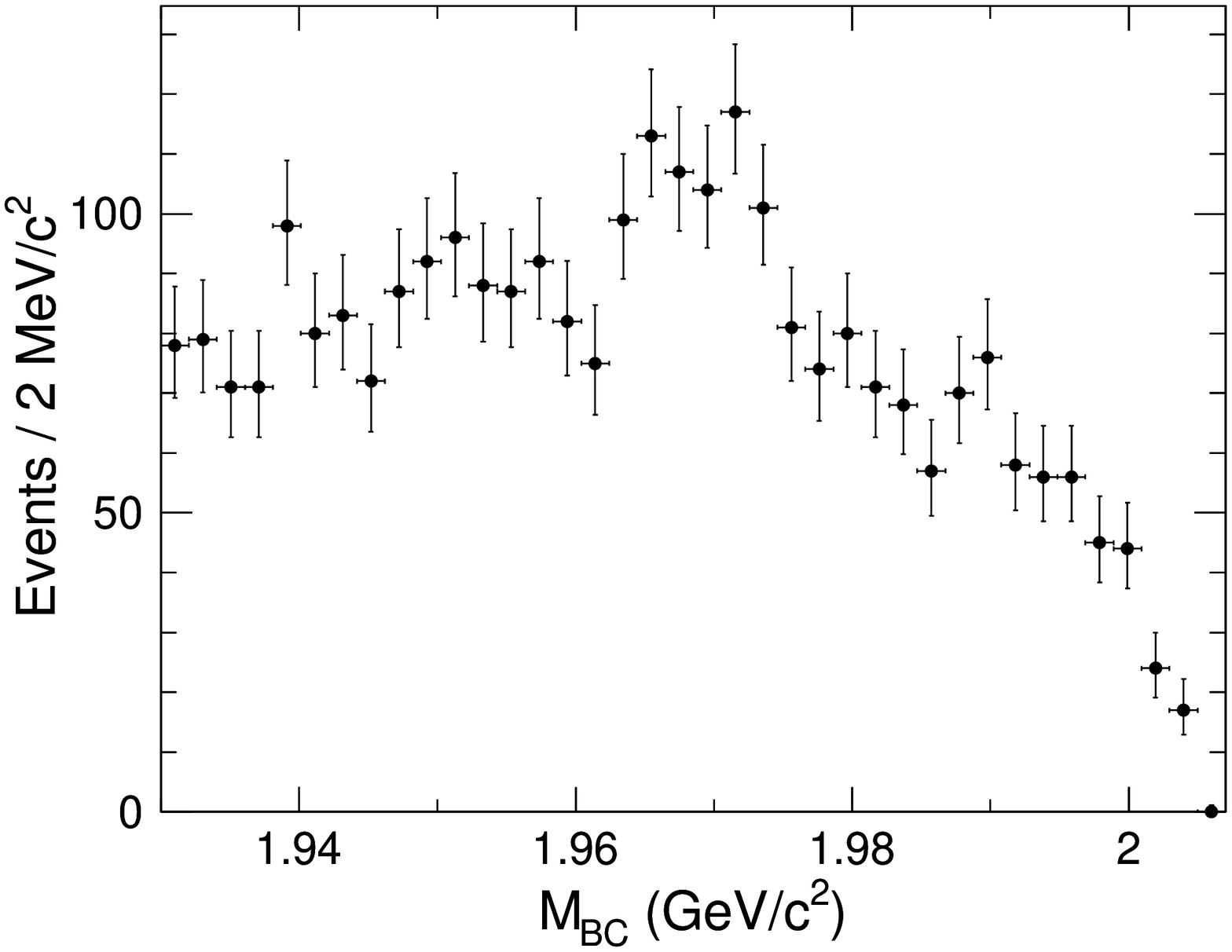}
\caption{$M_{\rm{BC}}$ distributions with the requirement of  $|\cos\theta_{\pi^{+}}|<0.5$ (left) or $|\cos\theta_{\pi^{+}}|>0.5$ (right).}
\label{fig:mBC-difctht}
\end{center}
\end{figure}

We perform a two dimensional unbinned maximum likelihood fit to the distribution of $M_{\rm{BC}}$ versus $\cos \theta_{\pi^{+}}$ to determine the yield $y^{\eta'\rho^{+}}_{\rm{ST}}$. The signal model of $M_{\rm{BC}}$ is the same as that in the analysis of $D_{s}^{+}\rightarrow\eta'X$. For $\cos\theta_{\pi^{+}}$, the signal shapes of $D_{s}^{+}\rightarrow\eta'\rho^{+}$  and $D_{s}^{+}\rightarrow\eta'\pi^{+}\pi^{0}$ are determined based on MC simulations. Background contributions in $M_{\rm{BC}}$ are modeled with an ARGUS function, while background contributions in $\cos\theta_{\pi^{+}}$ are taken from the events in the $M_{\rm{BC}}$ sidebands $1.932<M_{\rm{BC}}<1.950\gevcc$ and $1.988<M_{\rm{BC}}<1.997\gevcc$. There is no obvious correlation between $M_{\rm{BC}}$ and $\cos\theta_{\pi^{+}}$, so the PDFs used for these two variables are directly multiplied. Figure~\ref{fig:mbcVSctht} shows the projections of the two-dimensional fit results in data. In the right plot, we further require $1.955<M_{\rm{BC}}<1.985\gevcc$ to enrich signal events.
The fit returns $y^{\eta'\rho^{+}}_{\rm{ST}}=210\pm50$, and $y^{\eta'\pi^{+}\pi^{0}}_{\rm{ST}}=-13\pm56$, which indicates that no significant non-resonant $D_{s}^{+}\rightarrow\eta'\pi^{+}\pi^{0}$ signal is observed. An upper limit of $\mathcal{B}(D_{s}^{+}\rightarrow\eta'\pi^{+}\pi^{0})$ at the 90\% confidence level is evaluated to be 5.1\%, after a probability scan based on 2000 separate toy MC simulations, taking into account both the statistical and systematic uncertainties. 
As shown in Fig.~\ref{fig:mBC-difctht}, we see obvious $D_{s}^{+}$ signals in the $M_{\rm{BC}}$ distribution with the requirement of $|\cos\theta_{\pi^{+}}|>0.5$, while it is not the case when requiring $|\cos\theta_{\pi^{+}}|<0.5$. This indicates that the three body process is not significant.

We study the $M_{\rm{BC}}$ distributions for events in $\rho^+$ and $\eta'$ sidebands.
The $\rho^+$ sideband region is chosen as $M(\pi^{+}\pi^{0})<0.500\gevcc$, and the $\eta'$ sidebands are $0.915<M(\eta'_{\pi\pi\eta})<0.925\gevcc$ and $0.990<M(\eta'_{\pi\pi\eta})<1.000\gevcc$. No $D_{s}^{+}$ signal is visible in the sideband events, further substantiating that the non-resonant processes $D_{s}^{+}\rightarrow\eta'\pi^+\pi^0$ and $D_{s}^{+}\rightarrow\eta \pip\pim\rho^+$ are negligible. A simulation study shows that the potential background contribution from $\eta'\rightarrow \rho^{0}\gamma$ is negligible.

The detection efficiency $\varepsilon^{\eta'\rho^{+}}_{\rm{ST}}$ is estimated to be $(9.80\pm0.04)$\%. Combined with the results for the normalization mode $K^{+}K^{-}\pi^{+}$, as given in Table~\ref{table:STandDT}, we obtain from Eq.~\eqref{equ:eprho equ} the ratio of $\mathcal{B}(D_{s}^{+}\rightarrow\eta'\rho^{+})$ relative to $\mathcal{B}(D_{s}^{+}\rightarrow K^{+}K^{-}\pi^{+})$ as $1.04 \pm 0.25$. Taking the most precise measurement of $\mathcal{B}(D_{s}^{+}\rightarrow K^{+}K^{-}\pi^{+})$ = (5.55$\pm0.19$)\% from CLEO~\cite{cleo:newDs} as input, we obtain $\mathcal{B}(D_{s}^{+}\rightarrow\eta'\rho^{+})$ = (5.8$\pm$1.4)\%.

\subsection{Systematic uncertainties}

In the measurement of $\mathcal{B}(D_{s}^{+}\rightarrow \eta'X)$, many uncertainties on the ST side mostly cancel in the efficiency ratios in Eq.~\eqref{equ:sig equ}. Similarly, for $D_{s}^{+}\rightarrow \eta'\rho^{+}$, the uncertainty in the tracking efficiency cancels to a negligible level by taking the ratio to the normalization mode $D_{s}^{+}\rightarrow K^{+}K^{-}\pi^{+}$ in Eq.~\eqref{equ:eprho equ}. The following items, summarized in Table~\ref{table:sys err},  are taken into account as sources of systematic uncertainty.

\begin{table}[tp!]
\centering
\footnotesize
\caption{Summary of relative systematic uncertainties in percent. The total uncertainty is taken as the sum in quadrature of the individual contributions.}
\begin{tabular}{lcc}
\hline \hline
\multicolumn{1}{l}{Source}                         & $\mathcal{B}(D_{s}^{+}\rightarrow \eta'X)$  &  $\mathcal{B}(D_{s}^{+}\rightarrow \eta'\rho^{+})$\\
\hline
MDC track reconstruction                              &  2.0  &       \\
PID                                                   &  2.0  &  3.0  \\
$\pi^{0}$ detection                                   &       &  2.4  \\
$\eta$ detection                                      &  2.7  &  3.5  \\
$\Delta E$ requirement                                &  1.0  &  1.4  \\
$M(\eta'_{\pi\pi\eta})$ requirement                   &       &  2.0 \\
$M(\eta'_{\pi\pi\eta})$ backgrounds                    &  1.5  &      \\
Peaking backgrounds in ST                              &  0.3  &      \\
$M_{\rm{BC}}$ signal shape                            &  1.0  &  0.6 \\
$M_{\rm{BC}}$ fit range                               &  1.7  &  0.5 \\
$\cos\theta_{\pi^{+}}$ backgrounds                     &       &  2.9 \\
Uncertainty of efficiency                             &  1.6  &  0.5 \\
Quoted branching fractions                            &  1.7  &  3.8  \\
\hline
Total                                                 &  5.3  &  7.5 \\
\hline \hline
\end{tabular}
\label{table:sys err}
\end{table}

\begin{enumerate}[a.]
\item {\it MDC track reconstruction efficiency.}~~~The track reconstruction efficiency is studied using a control sample of $D^{+}\rightarrow K^{-}\pi^{+}\pi^{+}$ in the data sample taken at $\sqrt{s}$ = 3.773\,GeV. The difference in the track reconstruction efficiencies between data and MC is found to be 1.0\% per charged pion and kaon. Therefore, 2.0\% is taken as the systematic uncertainty of the MDC track reconstruction efficiency for $D_{s}^{+}\rightarrow \eta'X$.
\item {\it PID efficiency.}~~~We study the PID efficiencies using the same control sample as in the track reconstruction efficiency study. The difference in PID efficiencies between data and MC is determined to be 1.0\% per charged pion or kaon. Hence, 2.0\% (3.0\%) is taken as the systematic uncertainty of the PID efficiency for $D_{s}^{+}\rightarrow \eta'X$ ($D_{s}^{+}\rightarrow \eta'\rho^{+}$).
\item {\it $\pi^{0}$ and $\eta$ detection.}~~~The $\pi^{0}$ reconstruction efficiency, including the photon detection efficiency, is studied using a control sample of $D^{0}\rightarrow K^{-}\pi^{+}\pi^{0}$ in the data sample taken at $\sqrt{s}$ = 3.773\,GeV. After weighting the systematic uncertainty in the momentum spectra of $\pi^{0}$, 2.8\% is taken as the systematic uncertainty for the $\pi^{0}$ efficiency in $D_{s}^{+}\rightarrow \eta'\rho^{+}$.
    Similarly, the systematic uncertainty for the $\eta$ efficiency in $D_{s}^{+}\rightarrow \eta'X$ ($D_{s}^{+}\rightarrow \eta'\rho^{+}$) is determined to be 2.7\% (3.5\%) by assuming data-MC differences have the same momentum-dependent values as for $\pi^{0}$ detection.
    The systematic uncertainties were set conservatively using the central value of the data-MC disagreements plus 1.0 (1.64) standard deviations for $\pi^{0}$ ($\eta$), as appropriate for a 68\% (95\%) confidence level. Here we inflate the $\eta$ uncertainty, because the uncertainty of the $\eta$ detection is estimated referring to $\pi^{0}$.
\item {\it $\Delta E$ requirement.}~~~Differences in detector resolutions between data and MC may lead to a difference in the efficiencies of the $\Delta E$ requirements. In our standard  analysis procedure, we apply different $\Delta E$ requirements on data and MC, to reduce the systematic uncertainties.
    To be conservative, we examine the relative changes of the efficiencies by using the same $\Delta E$ requirements for MC as for data.
    We assign these changes, 1.0\% for $D_{s}^{+}\rightarrow \eta'X$ and 1.4\% for $D_{s}^{+}\rightarrow \eta'\rho^{+}$, as the systematic uncertainties on the $\Delta E$ requirement.
\item {\it $M(\eta'_{\pi\pi\eta})$ requirement.}~~~In the right plot in Fig.~\ref{fig:rho-ep-mass-comp}, the resolution of the $\eta'$ peak in MC is narrower than data.  We take the change in efficiency of 2.0\%, after using a Gaussian function to compensate for this resolution difference,
    as the systematic uncertainty of the $M(\eta'_{\pi\pi\eta})$ requirement for $D_{s}^{+}\rightarrow \eta'\rho^{+}$.
\item {\it $M(\eta'_{\pi\pi\eta})$ background contributions.}~~~In the measurement of $\mathcal{B}(D_{s}^{+}\rightarrow \eta'X)$, a two-dimensional fit is performed to the $M_{\rm{BC}}(\rm{ST})$ and $M(\eta'_{\pi\pi\eta})$ distributions. The uncertainty due to the description of the $M(\eta'_{\pi\pi\eta})$ background contributions is estimated by repeating the fit with higher order polynomial functions. We take the maximum relative change of 1.5\% in the signal yields as the systematic uncertainty on $M(\eta'_{\pi\pi\eta})$ background contributions.
\item {\it Peaking background contributions in ST.}~~~For the ST $D_{s}^{-}$ candidates, we study the potential peaking background contributions with the inclusive MC sample.
      We find that there is no peaking background contributions except for $D_{s}^{-}\rightarrow \pi^{+}\pi^{-}\pi^{-}$.
We consider the rate of peaking background contributions in the ST yields, and take 0.3\% as the systematic uncertainty of peaking background contributions in the ST events.
\item {\it $M_{BC}$ signal shape.}~~~To estimate the uncertainty in the $M_{BC}$ signal shape, we perform alternative fits with MC-determined signal shapes with different requirements on the truth matches. We take the resultant changes of 1.0\% and 0.6\% in $\mathcal{B}(D_{s}^{+}\rightarrow \eta'X)$ and $\mathcal{B}(D_{s}^{+}\rightarrow \eta'\rho^{+})$ as the systematic uncertainties, respectively.
\item {\it $M_{BC}$ fit range.}~~~We change the fit ranges of $M_{\rm{BC}}$ for ST modes, and take the resulting changes of 1.7\% and 0.5\% in $\mathcal{B}(D_{s}^{+}\rightarrow \eta'X)$ and $\mathcal{B}(D_{s}^{+}\rightarrow \eta'\rho^{+})$, as the systematic uncertainties, respectively.
\item {\it $\cos \theta_{\pi^{+}}$ background contributions.}~~~In the measurement of $\mathcal{B}(D_{s}^{+}\rightarrow \eta'\rho^{+})$, a two dimensional fit is performed to the $M_{\rm{BC}}$ and $\cos\theta_{\pi^{+}}$ distributions. The shape of the backgrounds in $\cos\theta_{\pi^{+}}$ is taken from the kernel-estimated distribution of the events in the $M_{\rm{BC}}$ sidebands with the kernel width parameter $\rho=2$ \cite{Cranmer:2000du}. The uncertainty due to the description of the $\cos\theta_{\pi^{+}}$ background contributions is estimated by repeating the fit with $\rho=1.5$. We take the relative change of 2.9\% in the signal yields as the systematic uncertainty on $\cos\theta_{\pi^{+}}$ background contributions.
\item {\it Uncertainty of efficiency.}~~~In the measurement of $\mathcal{B}(D_{s}^{+}\rightarrow \eta'X)$, we use the inclusive MC samples to determine $\varepsilon_{\rm{ST}}^{\alpha}$.
    The DT efficiency $\varepsilon_{\rm{DT}}^{\alpha}$ is determined by $\varepsilon_{\rm{DT}}^{\alpha} =\Sigma_{\beta}\mathcal{B}_{\beta}\varepsilon_{\rm{DT}_{\beta}}^{\alpha}/\Sigma_{\beta}\mathcal{B}_{\beta}$, where $\varepsilon_{\rm{DT}_{\beta}}^{\alpha}$ is obtained from MC simulated events of $D_{s}^{-}\rightarrow\alpha$ and $D_{s}^{+}\rightarrow\eta'\beta$, and $\beta$ refers to the five most dominant final states $\pi^{+}, K^{+}, \rho^{+}, e^{+}\nu_{e}$ and $\mu^{+}\nu_{\mu}$, and $\mathcal{B}_{\beta}$ is the decay rate of $D_{s}^{+}\rightarrow\eta'\beta$. We assign the world averages to the branching fractions of these five modes, except for $\mathcal{B}(D_s^{+}\rightarrow\eta'\rho^{+})$, which is taken from our measurement. The statistical uncertainties in $\varepsilon_{\rm{DT}_{\beta}}^{\alpha}$ and the $\mathcal{B}_{\beta}$ uncertainties are propagated to $\varepsilon_{\rm{DT}}^{\alpha}$. The uncertainties of $\varepsilon_{\rm{ST}}^{\alpha}$ and $\varepsilon_{\rm{DT}}^{\alpha}$ are propagated to $\mathcal{B}(D_{s}^{+}\rightarrow \eta'X)$ and yield a systematic uncertainty of 1.6\%. For the measurement of $\mathcal{B}(D_{s}^{+}\rightarrow \eta'\rho^{+})$, the uncertainty of the efficiency due to the limited MC statistics is estimated to be 0.5\%.
\item {\it Quoted branching fractions.}~~~The branching fractions of $\eta'\rightarrow \pi\pi\eta$, $\eta\rightarrow\gamma\gamma$, $\pi^{0}\rightarrow \gamma\gamma$ are taken from PDG~\cite{PDG}; the branching fraction for $D_{s}^{+}\rightarrow K^{+}K^{-}\pi^{+}$ is taken from CLEO's measurement~\cite{cleo:newDs}. Their uncertainties are 1.6\%, 0.5\%, 0.03\% and 3.4\%, respectively.
\end{enumerate}

\section{Summary and Discussion}
We measure the branching fraction $\mathcal{B}(D_{s}^{+}\rightarrow \eta'X) = (8.8 \pm 1.8 \pm 0.5)\%$, which is consistent with CLEO's measurement~\cite{Dobbs:2009aa}.
The weighted average of these two results is $\mathcal{B}(D_{s}^{+}\rightarrow \eta'X) = (10.3\pm1.3)\%$.
We also measure the ratio ${\mathcal{B}(D_{s}^{+}\rightarrow\eta'\rho^{+})} \,/\, {\mathcal{B}(D_{s}^{+}\rightarrow K^{+}K^{-}\pi^{+})} = 1.04 \pm 0.25 \pm 0.07$, from which we get $\mathcal{B}(D_{s}^{+}\rightarrow \eta'\rho^{+}) = (5.8\pm1.4\pm0.4)\%$. This is nearly half of CLEO's older result~\cite{CLEO1998}, but compatible with CLEO's newer measurement of $\br{D_{s}^{+}\rightarrow\eta'\pi^{+}\pi^{0}}$~\cite{cleo:newDs}, in which the resonant process $\eta'\rho^{+}$ is believed to dominate.
We also report a limit on the non-resonant branching ratio $\mathcal{B}(D_{s}^{+}\rightarrow \eta'\pi^+\pi^0)<5.1\%$ at the 90\% confidence level.
These results reconcile the tension between experimental data and theoretical calculation~\cite{Fusheng:2011tw}. Taking the world average values of other exclusive branching fractions involving $\eta'$ as input, we obtain the sum of exclusive branching fractions $\mathcal{B}(\dsplus\to\eta'K^{+}, \eta'\pi^{+}, \eta'\rho^{+}, \eta'l\nu_{l}) = (11.9\pm1.6)\%$, in which $l$ denotes $e^{+}$ or $\mu^{+}$, and where we have assumed that $\mathcal{B}(\dsplus\to\eta'\mu^{+}\nu_{\mu}) = \mathcal{B}(\dsplus\to\eta'e^{+}\nu_{e})$. This summed exclusive branching fraction is compatible with the new weighted inclusive result $\mathcal{B}(\dsplus\to\eta'X) = (10.3\pm1.3)\%$.

\section{Acknowledgments}
The BESIII collaboration thanks the staff of BEPCII and the IHEP
computing center for their strong support. This work is supported in
part by National Key Basic Research Program of China under Contract
No.~2015CB856700; National Natural Science Foundation of China (NSFC)
under Contracts Nos.~11125525, 11235011, 11275266, 11322544, 11335008, 11425524;
the Chinese Academy of Sciences (CAS) Large-Scale Scientific Facility
Program; the CAS Center for Excellence in Particle Physics (CCEPP);
the Collaborative Innovation Center for Particles and Interactions
(CICPI); Joint Large-Scale Scientific Facility Funds of the NSFC and
CAS under Contracts Nos.~11179007, 11079008, U1232201, U1332201; CAS under
Contracts Nos.~KJCX2-YW-N29, KJCX2-YW-N45; 100 Talents Program of CAS;
INPAC and Shanghai Key Laboratory for Particle Physics and Cosmology;
Program for New Century Excellent Talents in University ( NCET ) under Contract No. NCET-13-0342;
Shandong Natural Science Funds for Distinguished Young Scholar under Contract No. JQ201402;
German Research Foundation DFG under Contract No.~Collaborative
Research Center CRC-1044; Istituto Nazionale di Fisica Nucleare,
Italy; Ministry of Development of Turkey under Contract
No.~DPT2006K-120470; Russian Foundation for Basic Research under
Contract No.~14-07-91152; U.S.~Department of Energy under Contracts
Nos.~DE-FG02-04ER41291, DE-FG02-05ER41374, DE-FG02-94ER40823,
DESC0010118; U.S.~National Science Foundation; University of Groningen
(RuG) and the Helmholtzzentrum fuer Schwerionenforschung GmbH (GSI),
Darmstadt; WCU Program of National Research Foundation of Korea under
Contract No.~R32-2008-000-10155-0.

\bibliographystyle{model1a-num-names}



\end{document}